# Highly-scalable stochastic neuron based on Ovonic Threshold Switch (OTS) and its applications in Restricted Boltzmann Machine (RBM)

Seong-il Im[1,2], Hyejin Lee[3,4], Jaesang Lee[3,5], Jae-Seung Jeong[3,6], Joon Young Kwak[3], Keunsu Kim[4], Jeong Ho Cho[2], Hyunsu Ju[1*], Suyoun Lee[3,5*]

[1]*Center for Opto-Electronic Materials and Devices, Korea Institute of Science and Technology, Seoul 02792, Korea*

[2]*Department of Chemical and Biomolecular Engineering, Yonsei University, Seoul 03722, Korea*

[3]*Center for Neuromorphic Engineering, Korea Institute of Science and Technology, Seoul 02792, Korea*

[4]*Institute of Physics and Applied Physics, Yonsei University, Seoul 03722, Korea*

[5]*Department of Materials Science and Engineering, Seoul National University, Seoul 08826, Korea*

[6]*Division of Nano and Information Technology, University of Science and Technology of Korea, Daejeon 34113, Korea*

**Abstract**

Interest in Restricted Boltzmann Machine (RBM) is growing as a generative stochastic artificial neural network to implement a novel energy-efficient machine-learning (ML) technique. For a hardware implementation of the RBM, an essential building block is a reliable stochastic binary neuron device that generates random spikes following the Boltzmann distribution. Here, we propose a highly-scalable stochastic neuron device based on Ovonic Threshold Switch (OTS) which utilizes the random emission and capture process of traps as the source of stochasticity. The switching probability is well described by the Boltzmann distribution, which can be controlled by operating parameters. As a candidate for a true random number generator (TRNG), it passes 15 among the 16 tests of the National Institute of Standards and Technology (NIST) Statistical Test Suite (Special Publication 800-22). In addition, the recognition task of handwritten digits (MNIST) is demonstrated using a simulated RBM network consisting of the proposed device with a maximum recognition accuracy of 86.07 %. Furthermore, reconstruction of images is successfully demonstrated using images contaminated with noises, resulting in images with the noise removed. These results show the promising properties of OTS-based stochastic neuron devices for applications in RBM systems.

1. **Introduction**

Recently, the development of artificial intelligence (AI) systems has shown remarkable progress due to the advances in the machine-learning (ML) techniques, for example, the deep neural network (DNN)[1], spiking neural network (SNN)[2], etc. Nonetheless, since such software- and CMOS-based approaches are problematic in terms of energy efficiency and scalability, many efforts have been paid to develop novel energy-efficient ML techniques. Among them, the interest in the Restricted Boltzmann Machine (RBM)[3, 4, 5] is increasing as a generative stochastic artificial neural network because it beneficially utilizes a local learning rule, so-called the Contrastive Divergence[6, 7], which reduces the computational load and consequently, energy consumption remarkably. Additionally, the RBM can be used as a base unit of advanced generative models, Deep Belief Net (DBN)[8] and Generative Adversarial Networks (GAN)[9].

Essential to the hardware implementation of RBM are reliable stochastic neurons that generate random spikes and have the spike probability described by a sigmoidal function of operating parameters such as the strength of the input signal, external bias, etc. Since multiple Si-CMOS transistors are required to implement such a function, there is a growing interest in the development of dedicated hardwares based on novel materials and device structures, for example, magnetic tunnel junctions (MTJs)[10, 11] and resistive memory (ReRAM) devices[12, 13, 14]. Nevertheless, those devices are still problematic in building a large-scale RBM due to the complex device structure for MTJs and the complex operation scheme for ReRAM devices, respectively.

In this work, we propose a simple stochastic neuron device that is easy to fabricate and highly scalable with controllable switching probability, promising properties for RBM hardware implementation. It consists of an Ovonic Threshold Switch (OTS) and a resistor ($R_{load}$) connected in-series (1OTS+1R) as shown in Fig. 1a[15]. With a simple metal/amorphous chalcogenide/metal structure, the OTS features reversible electrical switching[16]. Furthermore, since it is well known that the switching delay time of the OTS is stochastic in nature[17], the interval between spikes is expected to be irregular, indicating that the suggested composite device may be a promising candidate for a true random number generator (TRNG). We have investigated the behaviors of the 1OTS+1R device for applications in RBM to find that it effectively works as a TRNG with the switching probability described by the Boltzmann distribution which is controllable by the width and the interval of input pulses. In addition, from a study using a simulated RBM neural network consisting of 1OTS+1R model stochastic neurons, the recognition and reconstruction of handwritten digits are successfully demonstrated, indicating that the proposed 1OTS+1R structure is promising for applications in hardware RBM.

## 2. Results

### Characteristics of OTS-based stochastic neuron and randomness test

Fig. 2a shows the representative switching behavior of the 1OTS+1R device with $R_L$=2kΩ, where $V_{out}$-waveform (voltage across $R_L$, red curve) shows the response of the device to consecutive pulses (black curve: width=150 ns, period=300 ns, $V_{in}$=2.1 V). It is found that the decay of $V_{out}$ is sometimes extended, leading to random "switching or not" behavior. Such an extended decay of $V_{out}$ means that the OFF state of the OTS device does not still recover after the bias is removed. Although the microscopic origin of the trap states in OTS devices remains controversial between the valence alternation pairs (VAPs) and other defects in amorphous chalcogenides[18, 19, 20, 21, 22, 23, 24, 25, 26], electrical switching in OTS devices is considered to arise from emission and capture of charge carriers from/to trap states for ON- and OFF-switching, respectively. In this regard, the extended decay of $V_{out}$ is attributed to the slower trapping process, which is stochastic in nature. Since the capture rate is known to be lower than the emission rate in OTS devices[27], the OFF-switching is more vulnerable to the stochastic nature consistent with the observation.

The region inside the red dashed box in Fig. 2a is enlarged in Fig. 2b. Comparing with the input pulses, notice that each output spike encloses different numbers of input pulses, clearly demonstrating the stochastic nature. This property can be exploited to build a TRNG. To investigate the feasibility, we use the scheme of periodic reading to assign a binary value at each point of reading depending on $V_{out}$ in comparison with a certain reference ($V_{ref}$ in Fig. 2b). As an example, the binary values are shown on the top of Fig. 2b using the scheme. To visualize the randomness of such a generated bitstream, a 70×70 black & white map is plotted using 4900 bits, confirming an irregular pattern (see Supplementary Fig. 1). In addition, using the National Institute of Standards and Technology (NIST) Statistical Test Suite (Special Publication 800-22)[28], we have performed the randomness test of the bitstream to find it passes 15 among the 16 tests (Table 1). The "Maurer's Universal Statistical Test" could not be tested due to the limit of the bitstream length and is believed to be satisfied with a much longer binary bit string of at least 387,840 bits generated by the highly reliable OTS devices.

Another requirement for applications in RBM is that stochastic neurons should show the switching probability ($P_{sw}$) following the Boltzmann distribution. Thus, we have investigated the $V_{in}$-dependence of $P_{sw}$ obtained from 200 consecutive input pulses with a fixed width and period at each $V_{in}$. We have also investigated the dependence of the distribution of $P_{sw}$ on the width and the period of the input pulse. Fig. 3a shows $P_{sw}$ as a function of the amplitude of $V_{in}$ at various pulse widths, keeping the pulse period at 300 ns along with the fitting curves by a sigmoidal function $P_{sw}(V_{in}; V_0, \Gamma) = 1/\left[1+\exp\left(-\frac{V_{in}-V_0}{\Gamma}\right)\right]$, where $V_0$ and $\Gamma$ are constants depending on the pulse width and the pulse interval. It clearly shows that all the $P_{sw}(V_{in}; V_0$ and $\Gamma)$ curves follow the Boltzmann distribution whose $V_0$ and $\Gamma$ increase with increasing the pulse width. The observed increase in $V_0$ and $\Gamma$ with the pulse width can be explained in terms of the reduction of the OFF-period, resulting in an additional delay in the decay of the spike as mentioned in Fig. 2a. In Fig. 3b and c, the dependence of $V_0$ and $\Gamma$ on the pulse width and period are displayed, respectively. These results demonstrate the controllability of $V_0$ and $\Gamma$ by changing operating conditions. In addition, we highlight that the 1OTS+1R device is controlled by electrical means enabling more precise and local control, unlike other stochastic neuron devices counting on the thermal[29] or magnetic[30, 31] means. Therefore, these results show promising characteristics of the 1OTS+1R device for applications in RBM.

### Simulation of an RBM: MNIST recognition and de-noising by reconstruction

The neural network used in this research consists of a hidden layer with 144 neurons and a visible layer with 794 neurons and, in the visible layer, there are 784 neurons corresponding to 28×28 pixel image data with 10 additional neurons for label data. The data set used for classification is an MNIST database consisting of 60,000 train data sets and 10,000 test data sets. Considering the cycle-to-cycle switching variation of the OTS device as shown in Fig. 3a,, the spiking probability variation at a specific voltage is described by the normal distribution at each voltage. The detailed RBM learning procedure is described in the Supplementary Information (see Supplementary Information S2).

Figs. 4a and b visualize the probabilities that change during the learning process in the visible and hidden layers of the OTS-based RBM. As the learning proceeds, the spiking probabilities of neurons in each layer are adjusted corresponding to the desired probability distribution. The probability distribution of the visible layer turns into a form similar to the input data as shown in Fig. 4a. On the other hand, the probability distribution of the hidden layer changes appropriately for the propagation of information to the visible layer by synaptic weights, shown in Fig. 4b. The spiking probabilities of neurons in each layer tend to become separately distributed by adjusting the intermediate probabilities toward 0 or 1 after the learning. The probability histogram of each layer is presented in the Supplementary Information (Supplementary Fig. 2), in order to illustrate the changes before and after the learning. Fig. 4c shows the classification accuracy on the test data set as every 1000 train data sets are trained. The classification accuracy of the OTS-based RBM is compared with one of the continuous probability distribution model (software) as the target reference accuracy. All models based on the device data show an accuracy of more than 85% and the maximum accuracy of 86.07% is obtained with a pulse width of 150 ns.

Fig.5a shows the classification accuracy of the OTS-based RBM on 10,000 test data with noise after training the same train data set with Figure 4(c). The test data with noise is obtained by including Additive White Gaussian Noise (AWGN)[32], of which the mean is 0 and the standard deviation is 0.1. Then the accuracy reaches over 81% in all the models based on the device data and the maximum accuracy is 83.08% with a pulse width of 125 ns. Even though the significant noise is added to the test data set, the classification accuracy remains at more than 80% because the RBM generates the imitative data with the training data set through the reconstruction process. Once the RBM learns the relation and the significance among the pixels of the training data set, the additive noise can be neglected or treated insignificantly through the reconstruction process. Therefore, the RBM model reproduces the image without noise even if noise exists in the input image. Fig. 5b and c show the input test images with the AWGN and the reconstructed images by the OTS-based RBM model, resulting in the de-noising effect[33].

3. Summary

In this research, we demonstrate that the OTS device successfully mimics a stochastic neuron required for RBM. The switching probability of the OTS is described by a sigmoidal function of the amplitude of $V_{in}$. The binary bit string generated by the OTS device passes the NIST tests for randomness, indicating that it is truly random. The classification in the MNIST data set is well demonstrated with the RBM based on the OTS device and the accuracy achieves 86.07%, even including the variation from run to run switching. Moreover, the classification accuracy on the MNIST data set with AWGN is 83.08% and the de-noising effect is observed through the trained OTS-based RBM. This OTS-based RBM device is advantageous to implement a stochastic neural network due to the simple structure of 1OTS+1R and thus provides a promising approach to overcome the scalability and the power consumption issues.

4. Methods

**Fabrication and characterization of 1OTS+1R devices**

Fig. 1b shows the vertical structure of an OTS device, which is fabricated with a pore-type structure. The pore size (*d*) is defined by the photolithography in the range of 2~20 μm. As a switching material, then 100 nm-thick GeSe layer is deposited in the pore by co-sputtering (magnetron RF sputtering) using $GeSe_2$ and Ge targets[34]. 100 nm- thick Pt and 50 nm-thick Mo layers are used as the bottom and top electrodes, respectively. The characteristic current-voltage (*I-V*) curve of an OTS device is shown in Fig. 1c, where 20 repetitions of measurements were made.

Device characteristics are investigated by using a conventional pulse-measurement setup[35], consisting of an arbitrary function generator (AFG-3101, Tektronix Inc.), a multi-channel oscilloscope (DPO-5104, Tektronix Inc.), and a source-measure unit (2635B, Keithley Inc.).

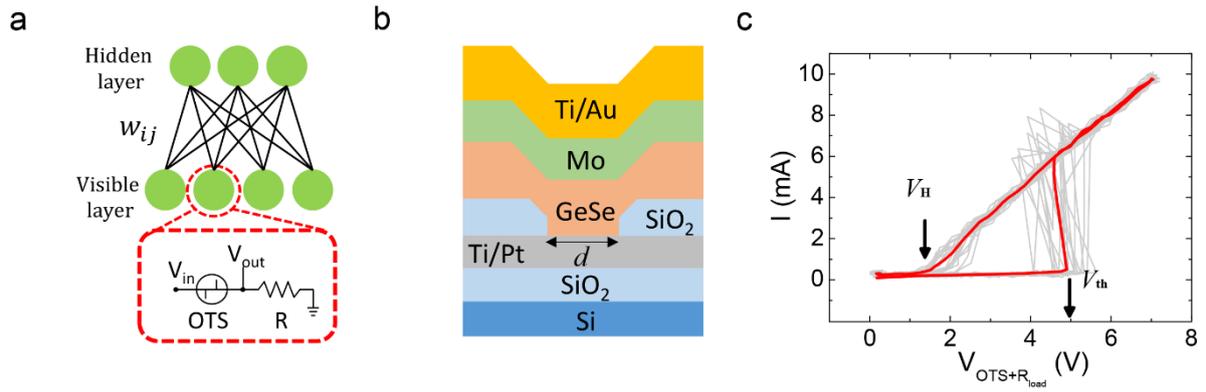

**Fig. 1 The structures of OTS and OTS based RBM. a** A conceptual RBM network consisting of 1OTS+1R stochastic neurons, **b** vertical structure of an OTS device, **c** characteristic *I-V* curve of an OTS device, where gray and red curves represent 20 repeated measurements and the averaged curve.

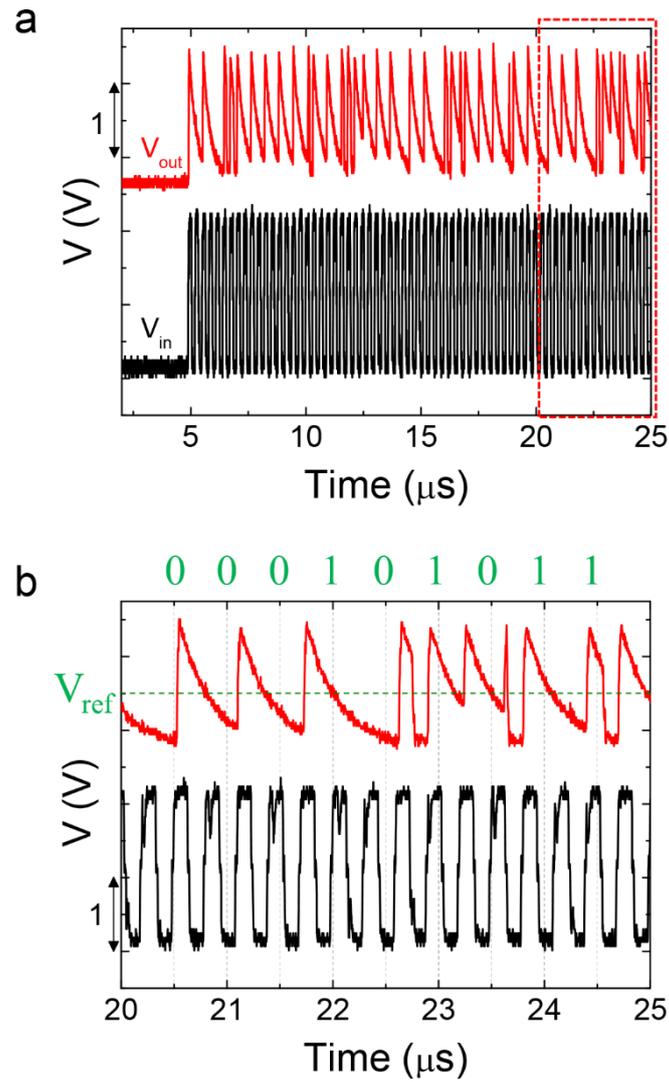

**Fig. 2 The characteristics of an OTS-based stochastic neuron. a** Representative $V_{in}$(black)- and $V_{out}$(red)-waveforms of the 1OTS+1R device $V_{in}$(black) is input as consecutive pulses (width=150 ns, period=300 ns) and $V_{out}$ waveform is vertically shifted for clarity. **b** Enlarged waveforms of the red box in (**a**). Vertical dotted lines represent points of reading to generate a binary bitstream. Some of the bitstream is shown on top of the graph as an example, which is obtained by comparing $V_{out}$ with $V_{ref}$.

| Type of Test | P-Value | Conclusion |
|---|---|---|
| 01. Frequency Test (Monobit) | 0.231 | Random |
| 02. Frequency Test within a Block | 0.018 | Random |
| 03. Run Test | 0.962 | Random |
| 04. Longest Run of Ones in a Block | 0.744 | Random |
| 05. Binary Matrix Rank Test | 0.540 | Random |
| 06. Discrete Fourier Transform (Spectral) Test | 0.012 | Random |
| 07. Non-Overlapping Template Matching Test | 0.045 | Random |
| 08. Overlapping Template Matching Test | 0.714 | Random |
| 09. Maurer's Universal Statistical test | -1.0 | Not available |
| 10. Linear Complexity Test | 0.130 | Random |
| 11. Serial test | 0.919 | Random |
| 11. Serial test | 0.906 | Random |
| 12. Approximate Entropy Test | 0.024 | Random |
| 13. Cummulative Sums (Forward) Test | 0.224 | Random |
| 14. Cummulative Sums (Reverse) Test | 0.021 | Random |
| 15. Random Excursion Test | 0.943 | Random |
| 16. Random Excursions Variant Test | 0.637 | Random |

**Table 1 Result of NIST statistical test for randomness.**

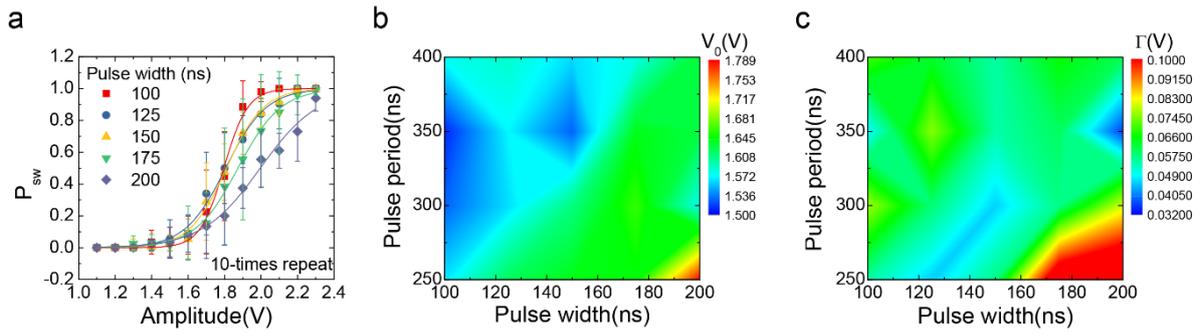

**Fig. 3 Controllable Boltzmann distribution. a** Switching probability ($P_{sw}$) as a function of the amplitude of $V_{in}$ with varying pulse widths with keeping the period at 300 ns. $P_{sw}$ is obtained from 200 samples at each condition and the errorbar is obtained from ten repeated measurements of $P_{sw}$. **b**, **c** Dependence of the mean ($V_0$) and the deviation ($\Gamma$) of the $P_{sw}$-distribution ($P_{sw}=1/(1+\exp[-(V_{in}-V_0)/\Gamma])$) on the pulse width and the period.

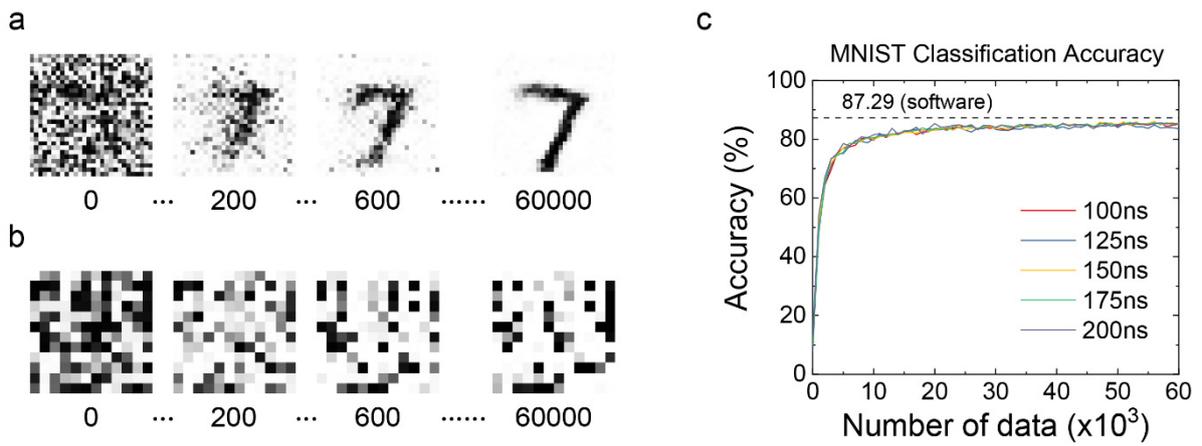

**Fig. 4 Changes in the spiking probability distributions of neurons. a** The visible layer and **b** The hidden layer. **c** Classification accuracy is evaluated with the switching probabilities with the pulse width from 100 ns to 200 ns for MNIST data set.

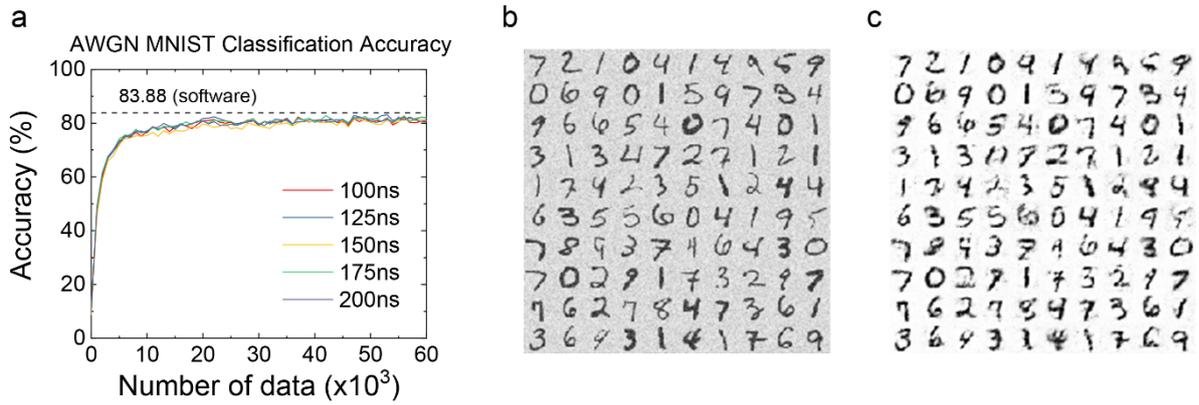

**Fig. 5 Reconstruction of images with de-noising effect. a** Classification accuracy is evaluated with the switching probabilities with the pulse width from 100 ns to 200 ns for AWGN MNIST test data set. **b** Examplary 100 MNIST test images with noise (AWGN) **c** and the reconstructed images by the trained OTS-based RBM.

## Acknowledgments

This work was supported by the Korea Institute of Science and Technology (KIST) through 2E30761 and 2E30100 and by National Research Foundation program through NRF-2019M3F3A1A02072175 and 2017R1E1A1A01077484.

## Author contributions

H.L. and S.I. equally contributed to this work. H.J. and S.L. conceived the concept. J.L. fabricated OTS devices. H.L. performed characterization of devices. S.I. performed the RBM simulation. J.J., J.K., K.K., and J.C. participated in the analysis of the randomness of the data. All authors discussed the results and participated in writing the manuscript.

## Competing interest

The authors declare no competing interests.

# Supplementary Information

# Highly-scalable stochastic neuron based on Ovonic Threshold Switch (OTS) and its applications in Restricted Boltzmann Machine (RBM)

Seong-il Im[1,2], Hyejin Lee[3,4], Jaesang Lee[3,5], Jae-Seung Jeong[3,6], Joon Young Kwak[3], Keunsu Kim[4], Jeong Ho Cho[2], Hyunsu Ju[1*], Suyoun Lee[3,5*]

S1.   Visualization of a generated bitstream (70×70 B&W map)

S2.   Details of RBM Learning

**S1. Visualization of a generated bitstream (70×70 B&W map)**

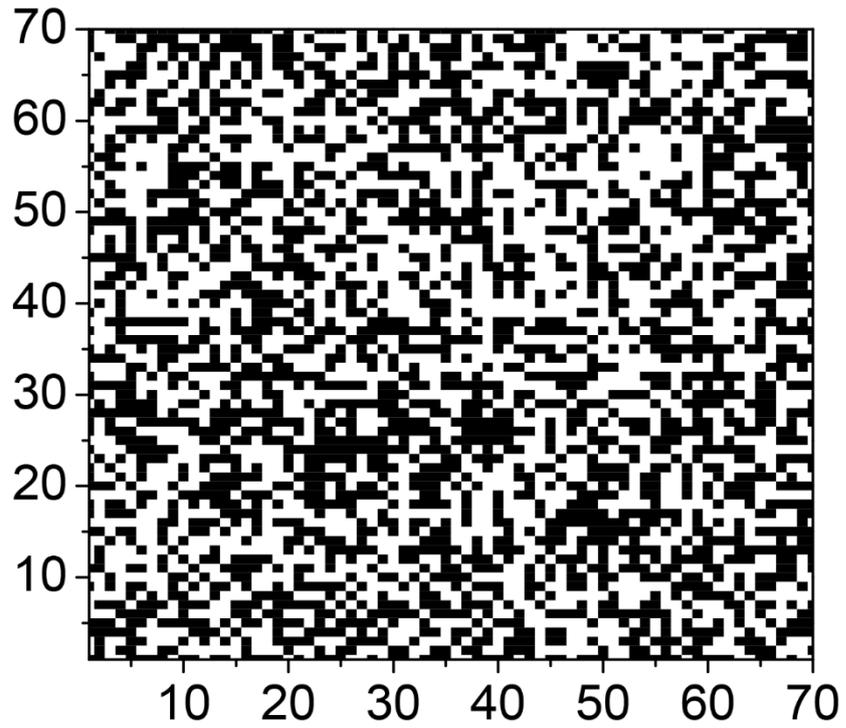

**Figure S1.** A 70×70 black & white map plotted using a generated bit stream (4.9 kbits). Black color represents "0" and white is "1".

## S2. Details of RBM Learning

The learning mechanism of RBM is divided into three parts: feed forward propagation, reconstruction, and parameter update. Feed forward propagation proceeds from the visible layer to the hidden layer. According to the probability distribution given as input data, the neurons in the visible layer are stochastically firing to have a binary value. The firing probability of neuron $h_j$ is calculated by Equation (1).

$$p(h_j = 1|v) = \frac{1}{1+e^{-(b_j+\Sigma_i v_i w_{ij})}} \quad (1)$$

The reconstruction process proceeds from the hidden layer to the visible layer. According to the probability obtained by the feed forward propagation, the neuron of the hidden layer has a binary value. Likewise, the neurons in the visible layer follow the firing probability described by Equation (2).

$$p(v_i = 1|h) = \frac{1}{1+e^{-(a_i+\Sigma_j h_j w_{ij})}} \quad (2)$$

Difference between the probability distributions of the input data and the reconstructed data can be optimized with the Kullback-Leibler Divergence(KLD)[1]. This method calculates an error between the two probability distributions and updates the probability parameters to reduce the error. To minimize the KLD, the gradient ascent method is used with log-likelihood as a cost function, given by

$$l = \log p(v)$$

$$\Delta w_{ij} = \eta \frac{\partial l}{\partial w_{ij}}$$

$$\frac{\partial l}{\partial w_{ij}} = \frac{\partial \log p(v)}{\partial w_{ij}} = <v_i h_j>_{data} - <v_i h_j>_{model}$$

$< v_i h_j >_{data}$ is the expectation value easily obtained from the input data, whereas $< v_i h_j >_{model}$ is evaluated through Gibbs Sampling[2], one of Markov Chain Monte Carlo Methods[3]. Through k-step contrastive divergence[4, 5], the Gibbs sampling is performed k-times and the last step results in $< v_i h_j >_{model}$. Then the probability parameter is correspondingly updated with the obtained $< v_i h_j >_{model}$.

**Probability distributions of each layer**

Figure S2(a) and (b) are the probability distribution graph showing the probability density of each layer before and after learning. Both figures tend to shift the spreading probabilities to 0, leaving only a few meaningful probabilities close to 1. In other words, learning proceeds by adjusting the spiking probability by the controlled voltage of each neuron. In the process of classification after learning, points representing 0 and 1 dominate, but in the process of learning, the model is affected by all points.

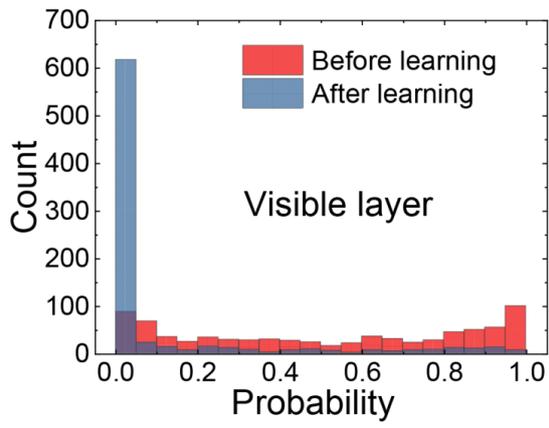 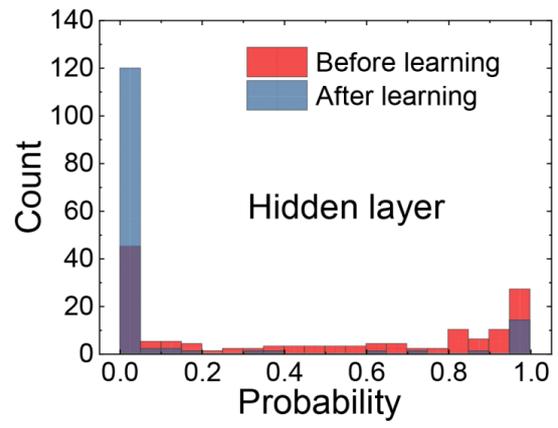

**Figure S2.** Probability histogram of the stochastic neurons in the hidden layer (a) and in the visible layer (b) before and after learning, respectively.